\documentclass[11pt]{scrartcl}

\usepackage[utf8]{inputenc}
\usepackage[english]{babel}
\usepackage{mathtools}
\usepackage{graphicx}
\usepackage{booktabs}
\usepackage{listings}
\usepackage{parskip}
\usepackage{subcaption}
\usepackage{xcolor}
\usepackage{hyperref}

\definecolor{psoe}{HTML}{E41A1C}
\definecolor{pp}{HTML}{377EB8}
\definecolor{podem}{HTML}{984EA3}
\definecolor{cs}{HTML}{FF7F00}
\newcommand{\psoe}[1]{\textcolor{psoe}{#1}}
\newcommand{\pp}[1]{\textcolor{pp}{#1}}
\newcommand{\podem}[1]{\textcolor{podem}{#1}}
\newcommand{\cs}[1]{\textcolor{cs}{#1}}

\begin{document}
\title{\textbf{Bayesian forecasting of electoral outcomes}\\ \textbf{with new parties' competition\footnote{We would like to thank David Rossell and two anonymous referees for their helpful comments.  This work was supported by the Spanish Ministry of Economy and Competitiveness [grant number ECO2017-82696P] and through the Severo Ochoa Program for Centres of Excellence in R\&D [grant number SEV-2015-0563] and the Government of Catalonia (ICREA-Academia, 2017SGR-616).}}}

\author{
  Garcia Montalvo, Jose\\
  \texttt{jose.garcia-montalvo@upf.edu}
  \and
  Papaspiliopoulos, Omiros\\
  \texttt{omiros.papaspiliopoulos@upf.edu}
  \and
  Stumpf-Fetizon, Timothee\\
  \texttt{tim.stumpf-fetizon@upf.edu}
}

\maketitle

\begin{abstract}
We propose a new methodology for predicting electoral results that combines a fundamental model and national polls within an evidence synthesis framework. Although novel, the methodology builds upon basic statistical structures, largely modern analysis of variance type models, and it is carried out in open-source software. The methodology is motivated by the specific challenges of forecasting elections with the participation of new political parties, which is becoming increasingly common in the post-2008 European panorama. Our methodology is also particularly useful for the allocation of parliamentary seats, since the vast majority of available opinion polls predict at national level whereas seats are allocated at local level. We illustrate the advantages of our approach relative to recent competing approaches using the 2015 Spanish Congressional Election. In general, the predictions of our model outperform the alternative specifications, including hybrid models that combine fundamental and polls' models. Our forecasts are, in relative terms, particularly accurate in predicting the seats obtained by each political party.

\end{abstract}

\section{Introduction}
\label{sec:intro}

Forecasting in social sciences is a challenging endeavour. Probably one of the most challenging exercises in this respect is the forecasting of election results. Most of the literature on election forecasting, including its methodological underpinning, has focused on two-party political systems, a "winner-take-all" system for the Electoral College and democracies with a long history of past elections. Conversely, in this paper we develop a methodology most appropriate for elections with little historical data for some competing parties, including the case of parties entering the electoral competition for the first time. Seats are allocated under a D'Hondt system. The vast majority of available opinion polls predict at national level whereas the seats are allocated at local level.

The scientific approach to electoral forecasting relies mostly on four alternative methodologies: the statistical modelling approach based on fundamentals; the use of polls, either voting intention surveys or party sympathy surveys; the use of political prediction markets based on bets for the candidates; and a combination of other methods, sometimes referred to as hybrid models\footnote{Recently there have been attempts to use social media to predict elections. Using Twitter has been found to be a poor forecasting strategy \cite{gayo}. \cite{murthy} shows that tweets are more reactive than predictive. \cite{wang} uses a Xbox gaming platform to show a new methodology to forecast elections in the context of highly non-representative survey data.}. The statistical modelling approach, also referred to as structural approach, consists of predicting election results from historical and socioeconomic data. An example is the simple "bread and peace" model of \cite{hibbs}\footnote{There are recent examples of fundamental models used in the context of Spanish political elections. For instance \cite{balaguer} show that an increase in local public spending increases the likelihood of being reelected.}. In stable political systems it is known that national election votes are highly predictable from fundamentals\footnote{For instance \cite{gelman1}.} while polls are highly variable but contain useful information, especially close to the election day. The aggregation of polls and the use of betting markets are further classical approaches to electoral forecasting. Recently there has been increasing interest in hybrid models which combine the outcomes of several methods. The most popular hybrid approach is the synthesis of a fundamental model and polls, e.g. \cite{lewis}. \cite{graefe} averages the results of pollsters, prediction markets, experts (journalists and scholars) and quantitative models to produce a combined forecast for the 2013 German election\footnote{See \cite{graefe2} for an application to the US elections.}. \cite{lewis2} and \cite{lewis3} present the canonical structure of this type of model. In this case the fundamental model is a regression on GDP and government popularity. This model is then synthesised with the median of polls, using a second regression, in order to predict the national level result.

Our methodology is also hybrid but it is tailored to situations where there is little historical data to apply existing hybrid methods. Moreover, elections are determined by seats won at local level, hence the national average is not that predictive of the party's representation in the parliament. A further reality particularly relevant to the European electoral landscape is that there is limited or no polling at local level. To put things in perspective, after the beginning of the financial crisis many new parties were created in European countries to capitalise on the discontent of voters with the policy reaction to the economic crisis. \cite{dennis} identify 45 "insurgent" parties in Europe, many of them just a few years old, that occupy the entire political spectrum, including the extremes. Insurgent parties held 1,329 seats in 27 EU countries in 2016, which correspond to 18.3\% of the total seats of their parliaments. The political landscape in Spain is complicated by the existence of numerous political parties with non-trivial representation in certain parts of the country (the so-called nationalist parties, e.g. in Catalonia or the Basque country), the fact that only a handful of elections have taken place since the restoration of democracy in the country in 1977 after decades of dictatorship, and that electoral polling is not as extensive as in older democracies (e.g. the USA or the UK). Moreover, as in most countries, polling is rarely available at higher spatial resolutions than national. However, by far the biggest challenge in the 2015 elections is that two new political parties ended up taking more than 30\% of the parliamentary seats even though they had no political representation in the previous parliament. Accordingly, the Spanish case is a good example of the challenges of forecasting electoral outcomes with new parties' competition.

Our approach is to learn the national average for each party primarily from published polls and use a fundamental model to learn how this national average propagates down to local level. In order to compensate for the lack of historical data for some competing parties, we use a fundamental model of voting intention \emph{trained} on "deep" micro-data obtained in the form of pre-electoral surveys. In Spain, these are carried out by the government-sponsored research centre CIS and allow us to estimate the relationship between geographical or demographic characteristics and voters' choice. The downside of these data sets is that the sample size in some provinces is very low and that the sample might not be representative. We address these issues through post-stratification\footnote{See Chapter 14 of \cite{gelmanbook}.} based on census data. This model is synthesised with a polls model, which computes weighted averages of published polls but, at the same time, corrects for potential sources of bias. These include house effects, the varying quality of polling methodologies, as well as time-trending that takes place as the election times approaches. Due to the lack of historical data the synthesis is not done by regression, but rather through a Bayesian evidence synthesis approach.

It is easiest to understand what that approach amounts to in the following way: the fundamental model produces simulations of local results for each party; these are transformed to local seats using D'Hondt method; the local results are aggregated at national level for each simulation; each of these simulations then receives a weight which corresponds to how close the implied national average is to the prediction of the polls model; then each implied national seat allocation is given the corresponding weight and weighted averages are computed to form predictions. We set up the fundamental model parameters so that the implied predictive distribution for the national average is fairly flat relative to that obtained from the polling model, hence the fundamental model is useful for learning how the national result propagates down to local level and for capturing correlations at that level.

Our approach has close links with recent work in election forecasting. Both the fundamental and the polling model are multilevel regression models.  \cite{park} use a multilevel regression model and post-stratification to obtain state level estimates from national polls\footnote{Multilevel structures are also relevant for some fundamental models. For instance \cite{elinder} shows that regional unemployment is a factor in the support for national governments.}. \cite{lock} use a Bayesian model to obtain a combination of polls with forecasts from fundamentals. They merge a prior distribution, obtained from previous election results, with polls to generate a posterior distribution over the result in each state relative to the national popular vote\footnote{For the national popular vote they use the model of \cite{hibbs}.}. The objective of this procedure is not to produce a forecast for the national vote but to develop a methodology that separates the national vote from states' relative positions. This can be very valuable for individual state forecasts.

The article is organised as follows. Section 2 introduces the challenge of forecasting electoral results in the presence of emerging parties and the political developments leading to the Congressional Elections of 2015. The choice of this example does not compromise the general applicability of our methodology: one of the challenges in forecasting electoral results is related to the fact that the allocation of seats may be very different from the proportion of votes at national level. Section 3 describes our methodology, starting with the fundamental model. It also describes the polls model and the synthesis of the two models. Section 4 applies the methodology proposed in Section 3 to the Spanish Congressional Election of 2015. Section 5 contains an evaluation of the forecasting accuracy of our model compared with some alternative models recently proposed. Finally, section 6 presents the conclusions.

\section{The Spanish 2015 Congressional Election}

Since the end of the dictatorship in 1977 Spanish politics was
characterised by the alternation in government of two political
parties: PP (popular party, conservative) and PSOE (socialists); see
Table 1 for the main contenders and their characteristics.
\begin{table}[htbp]
\centering
\begin{tabular*}{\textwidth}{@{\extracolsep{\fill} } lllr}
\toprule
Code & Party & Ideology & 2011 Result \\
\midrule
\textbf{\psoe{PSOE}} & Partido Socialista Obrero Espanol & Center-left & 0.288 \\
\textbf{\pp{PP}} & Partido Popular & Right-wing & 0.446 \\
\textbf{\podem{Pod}} & Podemos (including IU) & Left-wing & N/A \\
\textbf{\cs{C's}} & Ciudadanos & Center-right & N/A \\
\bottomrule
\end{tabular*}
\caption{Spanish parties active at national level in the 2015 elections.}
\end{table}

Some other small and regional parties also participated in the elections but the two largest parties accounted for 75\% to 85\% of the vote. The 2015 electoral campaign saw the emergence of two new contenders at national level: Podemos (radical left) and C's (Ciudadanos, liberal). Podemos and C's had no seats in previous Spanish parliaments\footnote{Podemos did not even exist at that time.}, whereas in the 2015 elections they ended up with 69 and 40 respectively, out of 350 in total. This structural change is one of the the main challenges in predicting the results of the 2015 Spanish Congressional Election through standard time series regressions. This challenge arises in any electoral contest where the emergence of new and large political parties change the electoral environment with respect to previous elections\footnote{Another challenging situation for electoral forecasting in the Spanish context arose in 2004 when a terrorist attack took place in Madrid during the last week before election day. No polls are allowed to be conducted during that period. See \cite{mont}. In recent years, terrorist attacks have frequently been timed to coincide with European elections. Obviously, the strategic timing of elections can also be triggered by good economic conditions or business cycle peaks. For a recent reference see \cite{canada}}.

The primary challenge from a modelling perspective is that Podemos and Ciudadanos have not inherited their electorate from a distinct previous political movement. On the contrary, they are cannibalising parties with similar ideologies. The following sections describe the modelling alternatives we considered and the difficulties that arise due to the emergence of these new political parties.

The dissatisfaction of a sizeable part of the population with the measures of austerity, initially introduced by the PSOE government in 2010, led to a popular demonstration that occupied the centre of Madrid during several weeks. This social movement was named 11-M since their assemblies had begun on May 11, 2011. On March 11, 2014 this movement crystallised into a new political party named Podemos, which soon had the support, in polls, of 15\% of the likely voters. Podemos was initially marketed as the Spanish Syriza\footnote{Syriza, or the Coalition of the Radical Left, is the Greek party that won the 2015 legislative election.}. The leaders of Podemos came mostly from Political Science university departments. Some of them had been members of leftist and anti-capitalist parties. While their program in the European elections of 2014 demanded the repudiation of public debt and the nationalisation of many industries, their positions later evolved in order to avoid the radical left label from their early days. Podemos ended up in coalition with IU (Izquierda Unida), the old communist party.

In addition, the conservative policies of PP, the corruption associated with conservative politicians and the lack of internal regeneration in the party led to the birth of a new liberal party called Ciudadanos (C's). This party was founded in 2006 but was initially geographically concentrated in Catalonia.

\begin{figure}[h!]
 \centering
 \includegraphics[width=0.80\textwidth]{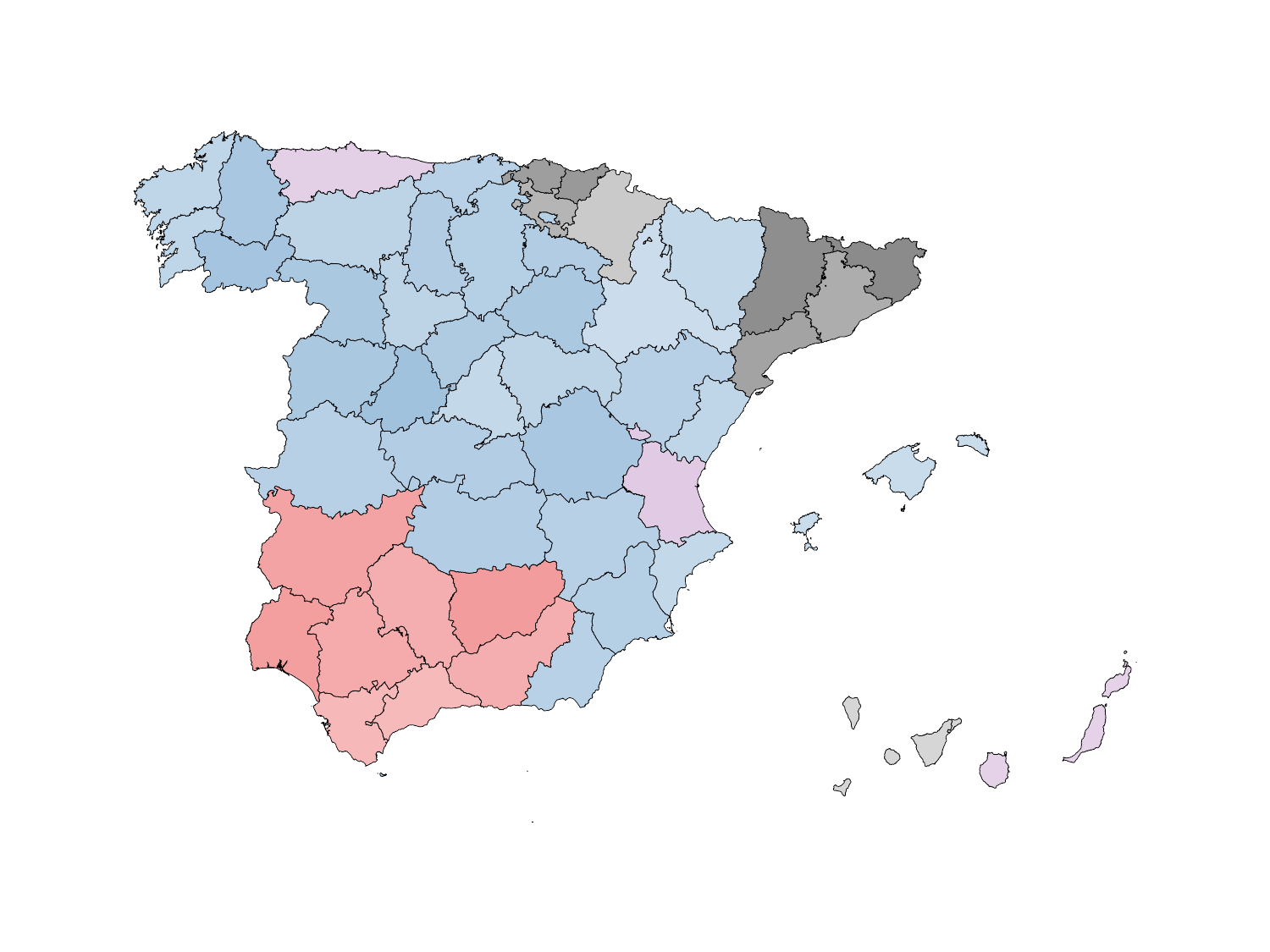}
 \caption{Map of Spanish provinces colored by strongest party in the
   2014 European elections and degree of dominance, darker shades corresponding to stronger dominance. Legend: PSOE (red), PP (blue), Podemos+IU (purple), others (gray).\label{fig:map}}
\end{figure}

Both Podemos and C's appear in the CIS\footnote{Center for Sociological Research (CIS) a publicly sponsored institution that runs the official polls. See Appendix A1 for the description of the data.} surveys as of July 2014. In contrast with Podemos, the support for Ciudadanos was only 0.9\% in early 2014 but built up quickly. In July 2015 polls showed a tie between these two new political contenders while the support for the two largest political parties had fallen to 50\%. Figure \ref{fig:map} depicts the strength of different parties in each province in the 2014 European elections. 

The Spanish government is appointed by the \emph{Congreso de los Diputados}, which consists of 350 representatives. Each of the 52 Spanish provinces elects its own representatives from its seat contingent according to the local electoral outcome. Thus, as in US presidential elections, the popular vote at national level is not decisive. Therefore, the notion of "local level" corresponds to "province level" in the Spanish electoral system, and we will use these two terms interchangeably in the rest of the article.

The allocation of seats at the province level is proportional, as opposed to the \emph{winner-takes-all} rule that most US states apply in presidential elections. The seat allocation is determined by the \emph{D'Hondt method} and is most easily understood in terms of the equivalent \emph{Jefferson method}, which we may express in terms of finding the market-clearing point in the market for seats\footnote{\cite{udina} use data on Spanish elections to show the forecast bias of pre-electoral polls when they convert votes into seats using D'Hondt's rule.}. The Jefferson method is used to find the ``price'' in votes per seat at which the ``demand'' for seats equals the available budget. Thus, a simple iterative algorithm consists of increasing the price per seat until the aggregated demand for seats equals the fixed supply. Then, each party obtains the number of seats it can afford at the equilibrium price. Since seats are an indivisible good, a party may just fall short of being able to buy an additional seat, with the remainder of its votes going to waste. This will occur in every province a party runs in. Thus, given a fixed national vote, it is preferable to have a geographically concentrated electorate. This applies to the nationalist parties in Catalonia and the Basque country.

In the Spanish case, there is an additional rule which states that parties must obtain at least 3\% of votes in a given province to take part in the allocation. Otherwise, their votes are disregarded. This acts as an additional penalty on smaller parties whose electorate is spread out across the nation.

\section{The proposed methodology}

This section provides a high-level description of the methodology we propose for predicting electoral outcomes in the presence of strong emerging parties. Firstly, we specify a fundamental model in the context of such parties. It predicts votes at local level from survey and census data. Secondly, we present a methodology for the aggregation of polls. It generates a cohesive forecast at national level. Finally, we propose a hybrid model that synthesises the other models. Estimation of these models, out of sample predictions for the 2015 Spanish elections, and comparisons - empirical and conceptual - with alternative fundamental, polls and hybrid models are deferred to Section \ref{sec:results}.

\subsection{Fundamental model with emerging parties}

The basic characteristics of our fundamental model are driven by the following considerations:

\begin{itemize}
\item It should generate predictions of votes at local level, from which we will then predict the seat allocation.
\item Since these local level predictions will be aggregated to the national level, it is statistically far more efficient (and less prone to biases) to aggregate probabilities computed at local level and turn them into point forecasts at national level, as opposed to providing point estimates at local level and then aggregating those. Effectively we are computing the expectation of a non-linear function of voting intentions at local level, and the exchange of function and expectation matters. Working with probabilities at local level also allows us to capture important correlations between outcomes at the different local units\footnote{See \cite{silver}}. We therefore adopt a probabilistic model of voting intention at local level, effectively a type of logistic regression.
\item Unlike in US Presidential Elections or previous Spanish Congressional Elections, voter choice is fundamentally not binary in the 2015 Congressional Election. Therefore, binary choice models are insufficient.
\item The drastic change of the political scene and the emergence of strong new parties renders historical models insufficient for prediction since there is little or no data to train them on.
\end{itemize}

To forecast the territorial distribution of votes we use data on individual respondents in pre-electoral surveys. In Spain, these are carried out by the government-sponsored research centre CIS\footnote{See Appendix A1.}. They allow us to estimate the relationship between geographical and demographic characteristics and voters' choice. The downside of these data sets is that the sample size in some provinces is small, leading to noisy estimates. Furthermore, their sample may be biased, and in any case our results should depend as little as possible on any potential bias due to non-representative sampling on the part of CIS.

We correct for both these issues by stratifying the respondents into disjoint "strata"; each "stratum" is a combination of different categorical variables, e.g. "man" (in the variable "sex"), in the age group 36-55 (in the variable "age group"), with tertiary schooling (in the variable "educational level"), employed (in the variable "employment status") who lives in a small community (in the variable "community size") in the province of Albacete (in the variable "province")". Say there are $N$ such strata (there are precisely 8424 in our specific application); and let $n$ be a specific stratum. We model the survey response counts $\boldsymbol{s}_{n}$ of a stratum $n$, which is the vector of counts in that stratum for the votes to each of the available parties, through a multinomial distribution:
\begin{gather}
  \boldsymbol{s}_{n} | \boldsymbol{\mu}_{n} \sim \operatorname{Multinomial}(\boldsymbol{\mu}_{n})
\end{gather}
where $\boldsymbol{\mu}_n$ is the vector of probabilities that a person belonging to such stratum votes for each of the available parties. From the most recent census data, we estimate $w_{i,n}$, the frequency of people in province $i$ that belong to stratum $n$; we then estimate the vector of probabilities of voting for each party in province $i$ as the weighted average
\begin{equation*}
\sum_{n=1}^N \boldsymbol{\mu}_n w_{i,n}.
\end{equation*}
This formula stems from the following basic decomposition:
\begin{align*}
& \textrm{Prob[vote party l $|$ province i]} = \\ & \sum_n \textrm{Prob[vote party l $|$ stratum n ] Prob[stratum n $|$ province i] }
\end{align*}
This approach is known as post-stratification, see e.g. \cite{park}.

The model we use for $\boldsymbol{\mu}_n$ is a multinomial logistic regression. In the case of two competing parties it becomes a logistic regression model for the probability of voting for one of the two parties given the stratum. For the multi-party setting we are interested in, let the vector $\boldsymbol{\mu}_n$ contain elements $\boldsymbol{\mu}_n(l)$, which is the probability of voting for party $l$, among  $l=1,\ldots,L$ competing parties. Then, the model becomes
\begin{equation*}
\boldsymbol{\mu}_n(l) = \frac{e^{f_{n,l}}}{\sum_{m=1}^L e^{f_{n,m}}}
\end{equation*}
where $f_{n,l}$ is a linear combination of dummy variables for the different levels of the categorical variables that define the stratum:
\begin{gather}
  f_{n,l} = \alpha_l + \sum_{k} \beta_{(k, j_{k}[n],l)}
\end{gather}
where $j_{k}[n]$ is the level of factor $k$ that corresponds to stratum $n$ for party $l$. The Appendix A2 and A3 contain details on the model and provide the multi-level formulae that define the model rigorously\footnote{We follow the standard practice of setting all coefficients of the pivot category (``other parties'') to 0 for simpler interpretation.}.

Therefore, we fit a main effects ANOVA model where each level of every categorical factor gets a different parameter. Additionally, we allow these parameters to differ for each party. The abundance of parameters calls for some type of regularisation, and we opt for a Bayesian multilevel approach, whereby the parameters associated with a factor are drawn from a common prior; see Appendix A4\footnote{\cite{steg} concludes that when using multilevel models the Bayesian approach is more robust and generates more conservative tests than the frequentist approach.}.

\subsection{Polls model}

\subsubsection{An explanatory ANOVA polls model}

Polls are published from a few months before until shortly before election day\footnote{In Spain a week before, but in Andorra it is allowed to publish polls regarding the Spanish elections up to a day before.} and give estimates of voting sentiment for each of the parties at national level. The simplest possibility to aggregate polls into a single prediction would be just to average the latest period (one week, two weeks, one month). This local averaging might be carried out using overlapping or non-overlapping windows of time. Forecasting can then be done only under the assumption that there is not going to be a change in public opinion from that time period to the election day.

This local averaging implicitly assumes that polls around a period in time are independent and identically distributed around the true voting sentiment. However, this is assumption is unlike to hold for a variety of reasons\footnote{In fact \cite{shirani}, in their analysis of 7,040 polls, show that there is a substantial election-level bias and excess of variance with respect to the variance calculated using the standard random sampling assumption.}:
\begin{itemize}
\item Polls by the same pollster may exhibit the same systematic bias across elections. For example, some pollsters are subject to political influence, which may lead them to bias their estimates systematically. This is known as the house effect\footnote{See \cite{silver} and \cite{shirani}}.
\item Polls preceding the same election may suffer from systematic bias across pollsters. This may be due to common methodological flaws and pollsters manipulating their polls to conform with the fold. We will call this an election effect\footnote{See \cite{silver}}.
\item Some pollsters' methodology may be superior, leading to lower error variance. Additionally, polls are carried out on samples of varying size\footnote{See \cite{shirani}}.
\item Subsequent polls may be trending up or down. We will refer to this as trending. \footnote{See \cite{lock} for evidence of trending close to election day and  \cite{linzer} for a stochastic trending model.}
\end{itemize}

We return to those later, after we have estimated our proposed model in section \ref{sec:results}, to show the evidence our data provide for each of those.

We can formalise these components. Let $\boldsymbol{p}_{k}$ denote a poll's prediction. Recall that in a multi-party system we have a vector of predictions, one for each competing party. Poll $k$ takes place at some time $t[k]$\footnote{This is standard multilevel notation. See Appendix A2 for details.}, and let  $\boldsymbol{v}_{t[k]}$ be the election result corresponding to poll $k$, i.e. the result of the election which this poll refers to. As earlier, let $\boldsymbol{p}_k(l)$ and $\boldsymbol{v}_{t[k]}(l)$ refer to the predictions and actual result for the $l$th party.
 We build a multi-level analysis of variance model for decomposing the error $\boldsymbol{p}_{k}(l)-\boldsymbol{v}_{t[k]}(l)$ as the sum of four terms:
\begin{itemize}
\item a time-invariant bias of the pollster that has produced the poll (house effect)
\item a pollster invariant bias that applies to each election separately (election effect)
\item a linear trend in time, with a coefficient that is allowed to vary across elections but is common to all pollsters (trending)
\item a poll-specific idiosyncratic error that could be due to differences in methodologies across pollsters and sampling variability.
\end{itemize}

Additionally, we learn the correlations between the idiosyncratic errors for different parties, and we allow the corresponding matrices to vary by pollster. Similarly, the effects that refer to different parties are allowed to be correlated, e.g. the house effects of a pollster for different parties. Again, the abundance of parameters calls for regularisation, and again we opt for a Bayesian approach to this multilevel model. All in all, the poll errors are modelled as a multivariate Gaussian distribution, the mean component and the covariance of which are implied by the decomposition described above\footnote{Polls of different pollsters in the same election are dependent through their dependence on the election effect, polls of the same pollster in different elections are dependent through their dependence on the common house effect, etc.}.
\begin{equation}
 (\boldsymbol{p}_{k} - \boldsymbol{v}_{t[k]}) \sim \operatorname{N}(\boldsymbol{\gamma}_{j[k]} + \boldsymbol{\delta}_{t[k]} + d_{k} \boldsymbol{\epsilon}_{t[k]}, \mathbf{\Sigma}_{j[k]})
\end{equation}
where $\boldsymbol{\gamma}_{j}$ is the time-invariant bias of pollster $j$, $\boldsymbol{\delta}_{t}$ is the pollster-invariant bias in election $t$, $d_{k}$ corresponds to how many days before the election poll $k$ was published and $\boldsymbol{\epsilon}_{t[k]}$ is the pollster-invariant strength of the trend in a given election. $\boldsymbol{\epsilon}_{t[k]}$ decays as election day approaches, but $\boldsymbol{\delta}_{t}$ applies to all polls until the election\footnote{See also \cite{montalvo2}. The model in \cite{shirani} includes a bias for each poll that is allowed to change linearly over time and a variance term that captures residual variability.}. As with the fundamental model we use the Bayesian multilevel paradigm to deal with the abundance of parameters and refer to Appendix A4 for details on the prior distributions we have used.

In summary, we build an ANOVA model to \emph{explain} the errors $\boldsymbol{p}_k(l)-\boldsymbol{v}_{t[k]}(l)$. From this perspective, this is not a predictive model. Its purpose is to understand the importance and relative magnitude of different sources of variability in published polls.

\subsubsection{Turning the explanatory into a predictive polls model}
\label{ssec:predict_poll}

The model we propose in the previous section implies a joint density for all the available polls in a given election conditional on the election result:
\[
\textrm{Prob[available polls $|$ new election result]}
\]
This density is obtained through the linear transformation
\[
\textrm{poll = result + poll error}
\]
and the model for the poll error we have built already. However, what we need is the "inverse probability"
\[
\textrm{Prob[new election result $|$ available polls ]}
\]
which we obtain by applying Bayes' theorem, as
\begin{align*}
& \textrm{Prob[new election result $|$ available polls ]} \propto \\
& \textrm{Prob[available polls $|$ new election result] Prob[new election result]}.
\end{align*}
Therefore, to get a predictive model we need a prior model to be combined with the explanatory model we have built. The approach followed here is an instance of what is known as \emph{inverse regression}, a popular approach to predictive modelling with high-dimensional data\footnote{See \cite{cook}}. In the hybrid model we propose in the sequel we get the fundamental model to serve as a prior. A simpler alternative, which is good enough for the purpose of predicting national average results but not seat allocation, is to use a uniform prior on the result, in which case
\begin{align*}
& \textrm{Prob[new election result $|$ available polls ]} \propto \\
& \textrm{Prob[available polls $|$ new election result]},
\end{align*}
the latter seen as a function of "new election result". Effectively, we exploit the symmetry of the Gaussian distribution in our errors model to create the predictive model as:
\[
\textrm{result = poll + poll error}
\]
The details on how to generate predictions using this predictive model are included in the Appendix A4.

\subsection{The hybrid model}

The basis of our hybrid model is the conditional probability we obtained in the previous section:
\begin{align*}
& \textrm{Prob[new election result $|$ available polls ]} \propto \\
& \textrm{Prob[available polls $|$ new election result] Prob[new election result]}.
\end{align*}
We use the explanatory polls model to produce the first density and the fundamental model based on surveys to produce the second. Operationally, we carry out the following procedure:
\begin{enumerate}
\item We produce a simulations of local results according to the fundamental model; let one such simulation be $\boldsymbol{v}_{i,s}$ for $i=1,\ldots,I$, where $i$ indicates local district and $s$ the simulation count.
\item For each simulation we aggregate result at national level to obtain a simulation of $\boldsymbol{v}_{s}$.
\item Provide weight to each simulation according to
\begin{equation*}
\mathcal{W}_s = \textrm{Prob[available polls $|$ $\boldsymbol{v}_{s}$ ]}
\end{equation*}
which is computed from the explanatory polls model.
\item Produce predictions by computing weighted averages
\begin{equation*}
\frac{\sum_{s=1}^S g(\boldsymbol{v}_{1,s},\ldots,\boldsymbol{v}_{I,s}) \mathcal{W}_s}{\sum_{s=1}^S \mathcal{W}_s}
\end{equation*}
\end{enumerate}
where $g$ is a function of interest of local results. We are particularly interested in the function that, implied by D'Hondt's method, maps local level results to national level seat allocations for each party. Apart from point estimates we can also produce predictive intervals in a similar way.

In the implementation described above one wants to make $S$, the total number of simulations, as large as possible, the limiting factor is the computational budget described in Section \ref{ssec:predict_poll}. In Section \ref{sec:results} we show graphical and numerical evidence that the resulting inference on the national-level result is dominated by the polls model; actually we devise a metric according to which we obtain that less than 35\% of the hybrid model inference is driven by the prior. In the appendix we explain how we get the fundamental model to have minor influence on the hybrid prediction of national averages.

\subsection{The computational approach underpinning the methodology}

The fundamental model and the polls model are ANOVA-type models, formulated as Bayesian multilevel models as described in some detail in the appendix A3 and A4. We follow a fairly standard specification of the prior distributions for these models, as for example explained in \cite{gelman2, gelmanbook}. We implement these models in \emph{Stan} (\cite{stan}), an open-source platform for performing full Bayesian statistical inference, which carries out Hamiltonian Markov chain Monte Carlo sampling from the corresponding posterior distributions.

Behind the scenes of the procedure described for the hybrid model, we use \emph{importance sampling}\footnote{See Chapter 2 of \cite{liu}.} to explore the distribution
\begin{equation*}
\textrm{Prob[new election result $|$ available polls \& survey data ],}
\end{equation*}
and a sampling-based approach to carry out the Bayesian updating\footnote{See \cite{smith}}.

\section{Results for the 2015 Spanish congressional election}
\label{sec:results}

\begin{figure}[b!]
\centering
\includegraphics[width=0.48\textwidth]{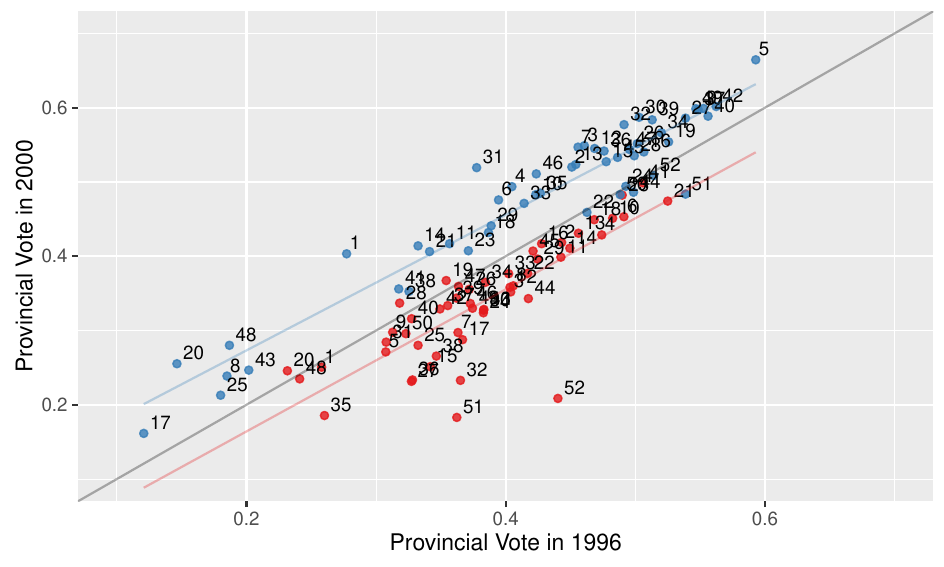}
\includegraphics[width=0.48\textwidth]{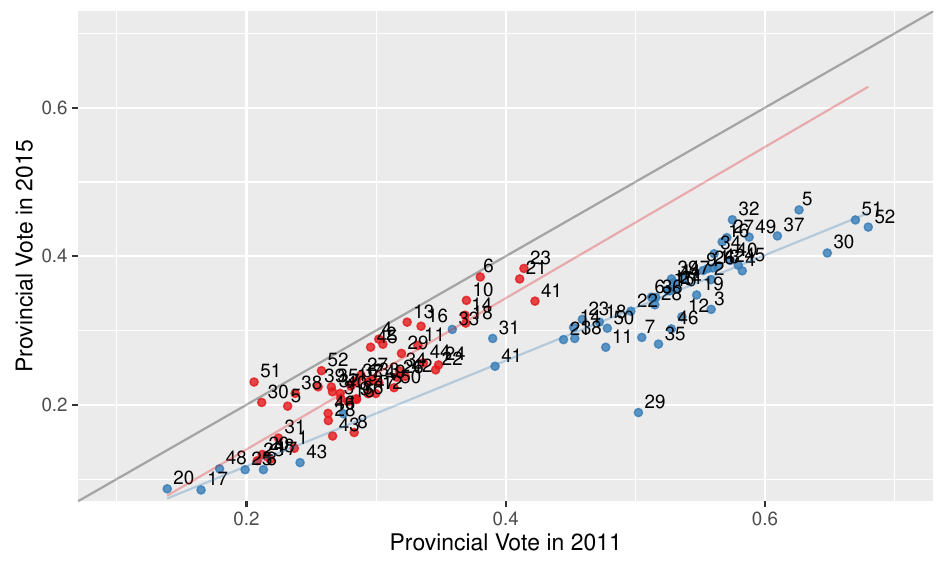}
\caption{Scatterplot of lagged vote share vs current vote share in
  2000 (left) and 2015 (right) relative to previous result, plus
  robust linear regression line. Legend: PSOE (red), PP (blue). In
  grey the $45^o$ line. The labels refer to the INE province code. \label{fig:autoreg}}
\end{figure}

In this section we apply the methodology discussed previously to the case of the Spanish Congressional Election of 2015. Before 2015, the national outcome was highly predictable from fundamentals and past results. For example, Figure \ref{fig:autoreg} (left) plots the electoral result of the 2000 election versus that of the 1996 election for the PP (blue) and the PSOE (red) in each province of Spain, numbered according to the standard postcode coding of Spanish provinces. The picture is similar in other elections prior to 2015. In particular, the positions of provinces relative to the national average are highly predictable. This manifests itself through regression lines that are almost parallel the $45^o$ line. Historical models typically include predictors such as unemployment rates, growth of personal disposable income, lagged electoral outcomes, presidential incumbency, regional trends, presidential approval, presidential home advantage (or the corresponding adjustment for party leader home province), partisanship or ideology of each state/district, etc.\footnote{See, for instance, \cite{camp2,camp1}, \cite{gelman1}, \cite{klarner}, \cite{fair} or \cite{hummel}.}

However, the usefulness and predictive ability of historical models for the 2015 Congressional Election is questionable. To start with, such a model cannot provide predictions for parties with no competitive participation in previous elections. Additionally, even when making predictions for the traditional political players in Spain, it is unlikely that the model estimated under a completely different political environment would have any applicability in this new situation. Figure \ref{fig:autoreg} (right) shows that the pattern observed in past elections has indeed been broken in the 2015 elections.

\subsection{Learning the fundamental model}
\label{sec:spanish-fund}

\begin{figure}[htbp]
 \centering
 \includegraphics[width=\textwidth]{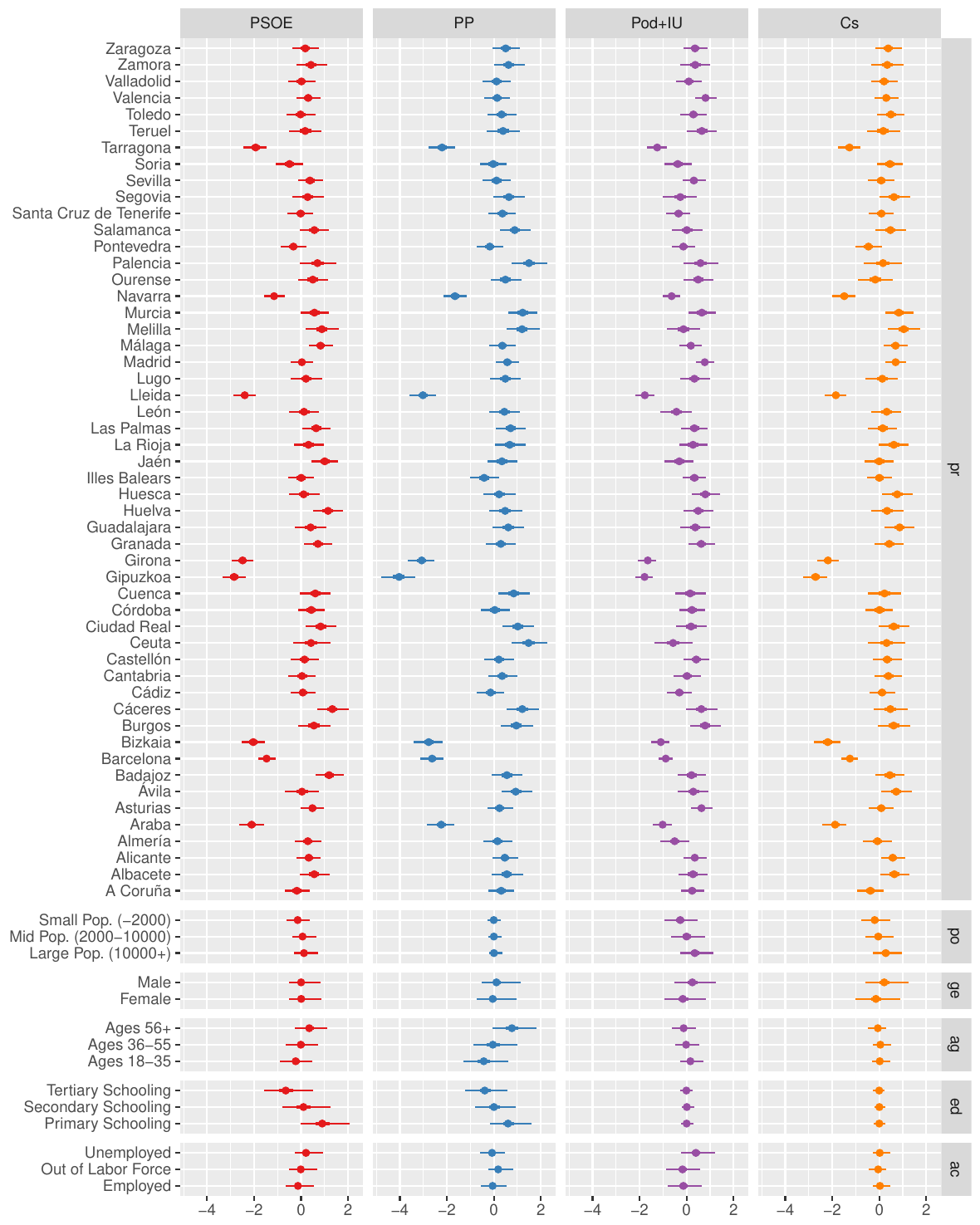}
 \caption{$\boldsymbol{\beta}_{j_{l}}$ marginal distributions (sentiment model level coefficients): median (point), 50 percent credibility interval (thick line) and 95 percent credibility interval (thin line). \label{fig:sentiments}}
\end{figure}

To train the fundamental model for the 2015 Spanish Congressional Election, we use the 2015 CIS pre-electoral survey\footnote{These surveys already contain questions about the two new parties.}. This results in a sample of about 17,452 respondents. We drop all respondents that did not report their voting intention from the sample, which amounts to assuming that their voting intention data is \emph{missing at random}. We include factors for the province, size of the municipality, gender, age, education and labour market activity of the respondent. The categories of each variables are described in Table \ref{tb:factors} of Appendix A1. The results of the estimation of the Bayesian multinomial logit model are summarised in Figure \ref{fig:sentiments}.\footnote{We run 4 chains in \emph{Stan} with 2000 iterations each, the first half of which we discard.}.

In practice, because responses may not accurately reflect behaviour at the voting booth, and because of the possibility of shifts between the survey and the election date, we inflate the uncertainty about the constants $\boldsymbol{\alpha}$, reflecting uncertainty about the national vote. In practice, this corresponds to multiplying MCMC draws of $\boldsymbol{\alpha}$ by 1.5\footnote{The choice of that factor is partially motivated by computational constraints that arise when combining the fundamental model with the polls.}.

To interpret the estimates, notice that:
\begin{itemize}
\item Since the model is overparameterised, some parameters are weakly identified. This manifests in wide marginal distributions. These overstate uncertainty since the parameters are highly correlated: once one of the parameters has been fixed, the uncertainty resolves.
\item As we fix the parameters of the pivot party (essentially consisting of regional parties) to 0, all estimates must be interpreted \emph{relative} to these parties. Therefore, a positive intercept estimate for PSOE implies that the average respondent is more likely to vote for PSOE than regional parties.
\end{itemize}

The province effects in Figure \ref{fig:sentiments} indicate that PP has the most variable territorial distribution, while Podemos is fairly constant. PSOE is strong in Andalucia and Extremadura and fairly weak in Catalonia and the Basque Country. PP has its strongest base in Castile and Murcia, but is extremely weak in Catalonia and the Basque Country.

\begin{figure}
\centering
\includegraphics[width=\textwidth]{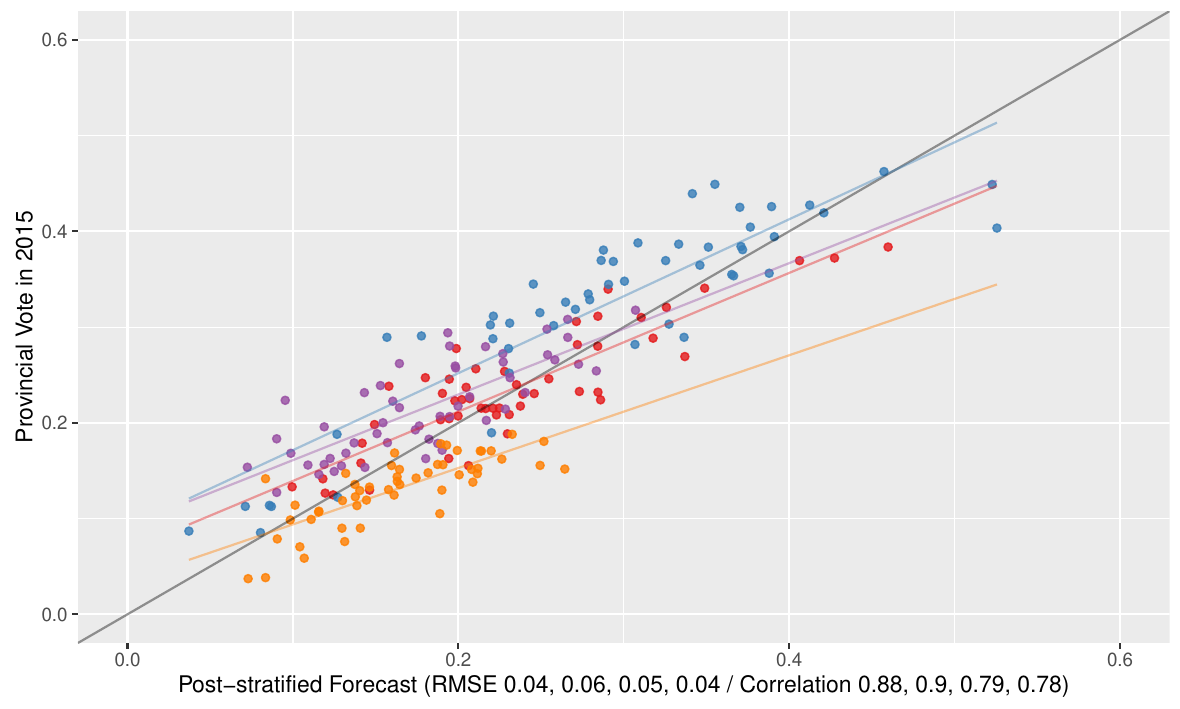}
\caption{Scatterplot of post-stratified point estimates vs. outcomes and regression line. Statistics are listed in the usual party order. MSE is computed as the average squared difference between the mean prediction and the result over provinces. Legend: PSOE (red), PP (blue), Podemos+IU (purple), C's (orange). \label{fig:poststrat}}
\end{figure}

As to the other factors, Podemos and C's are slightly more urban while the other parties' support does not vary along that dimension. PSOE and PP mostly appeal to uneducated voters. Labour market activity is likely irrelevant after controlling for other factors. Figure \ref{fig:poststrat} illustrates point predictions of the fundamental model after post-stratification and how they compare against the actual 2015 election results.

\subsection{Learning the pollster model}
\label{sec:spanish-poll}

We use the pollster model we have described in the 2015 Spanish
elections. We work with the results of 157 electoral polls published
before the Congressional Elections of 1996, 2000, 2004, 2008 and
2011. This set corresponds to the subset of polls published up to 30
days before a Congressional Election.

\begin{figure}
 \centering
 \includegraphics[width=\textwidth]{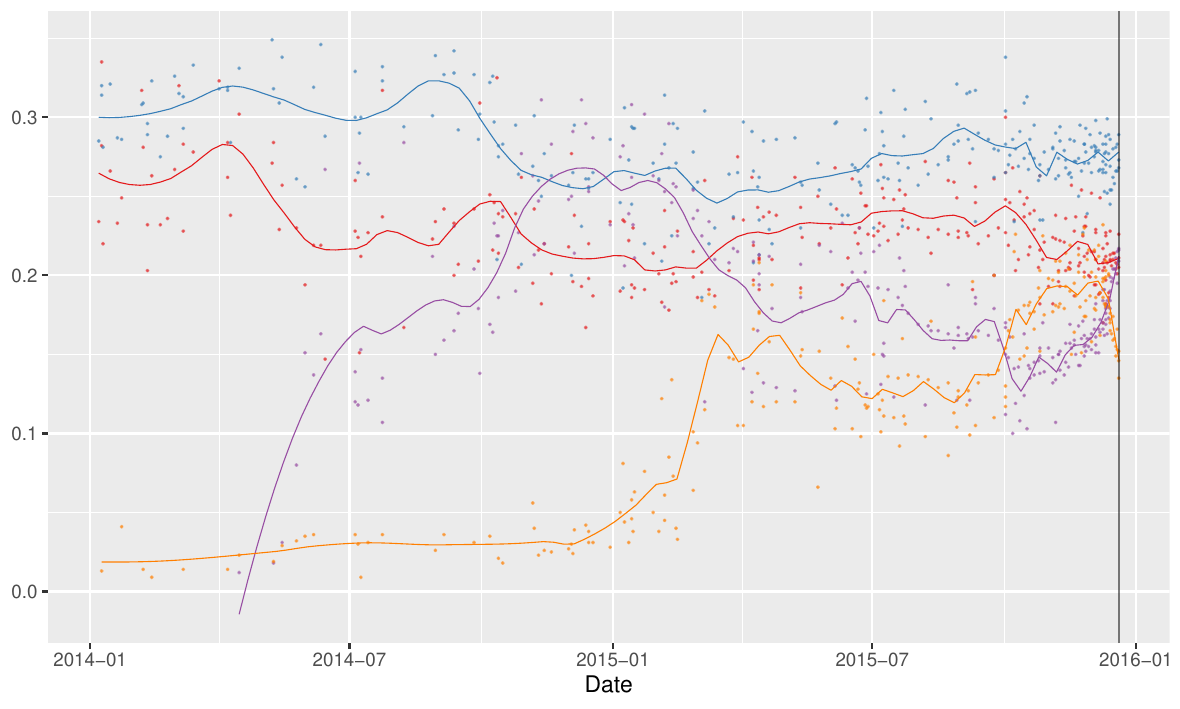}
 \caption{Polling before the general election of 2015, with LOESS
   smoother. Legend: Legend: PSOE (red), PP (blue), Podemos (purple),
   C's (orange). The election day is marked by the vertical solid line.\label{fig:smooth}}
\end{figure}

\begin{figure}[htbp]
 \centering
 \includegraphics[width=\textwidth]{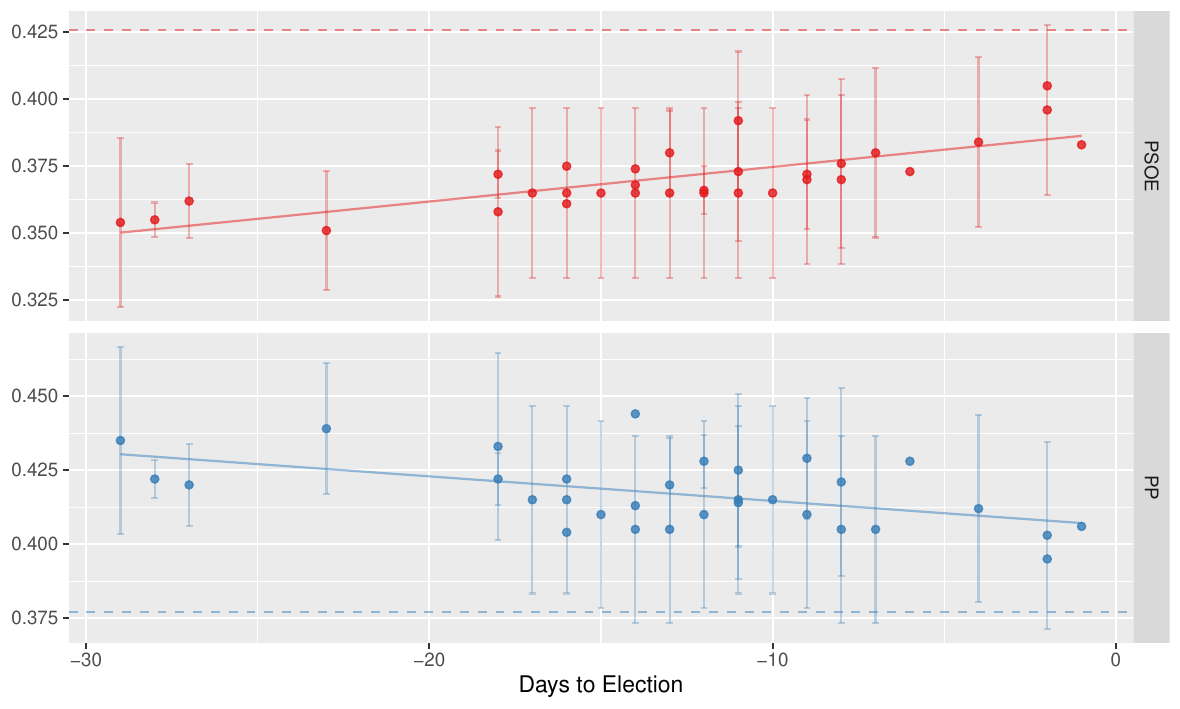}
 \caption{Polling before the general election of 2004. The solid line is a linear trend (OLS), the dashed line is the election result and the error bars correspond to the margin of error reported by the pollster. \label{fig:err}}
\end{figure}

Exploratory analysis reveals that the uncertainty about the election
result close to election day by far exceeds the sampling
uncertainty\footnote{This result is also supported by the evidence provided in \cite{shirani}.}. Rolling averages, like the ones depicted in Figure \ref{fig:smooth}, do not provide a direct measure of uncertainty, which is essential to building a probabilistic model. Averaging multiple polls does not eliminate the excess uncertainty. Furthermore, we sometimes observe sharp trending close to election day, even after prolonged periods of stability. Following (and extrapolating) the trend usually takes us closer to the election result than simple averaging. Figure \ref{fig:err} shows for the 2015 Spanish elections how the declared margin of error in the polls, usually given as the inverse of the square root of the sample size, tends to underestimate the true uncertainty. Furthermore, using a linear trend brings us closer to the election result.

\begin{figure}
 \centering
 \includegraphics[width=\textwidth]{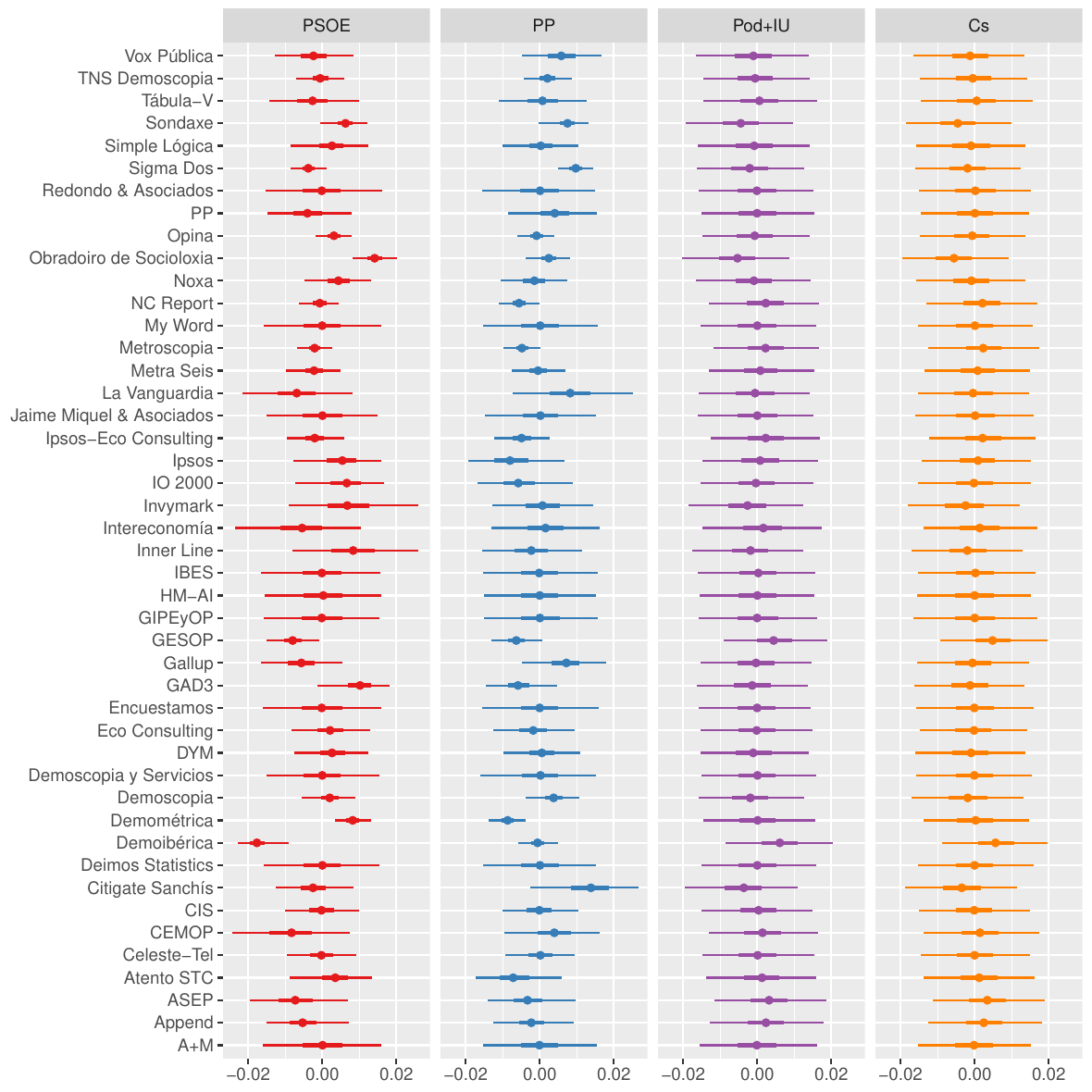}
 \caption{$\boldsymbol{\gamma}_{j}$ marginal distributions (pollster bias): median (point), 50 percent credibility interval (thick line) and 95 percent credibility interval (thin line). Positive values imply that the pollster is overestimating. \label{fig:polls_bias}}
\end{figure}

\begin{figure}
 \centering
 \includegraphics[width=\textwidth]{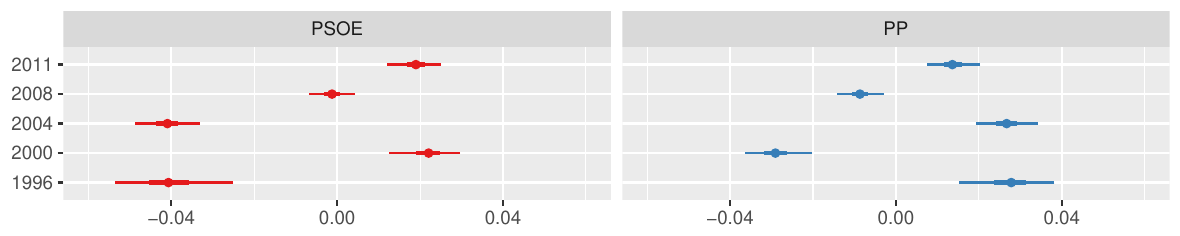}
 \caption{$\boldsymbol{\delta}_{t}$ marginal distributions (election bias): median (point), 50 percent credibility interval (thick line) and 95 percent credibility interval (thin line). Positive values imply that the pollster is overestimating. \label{fig:election_effect}}
\end{figure}

\begin{figure}
 \centering
 \includegraphics[width=\textwidth]{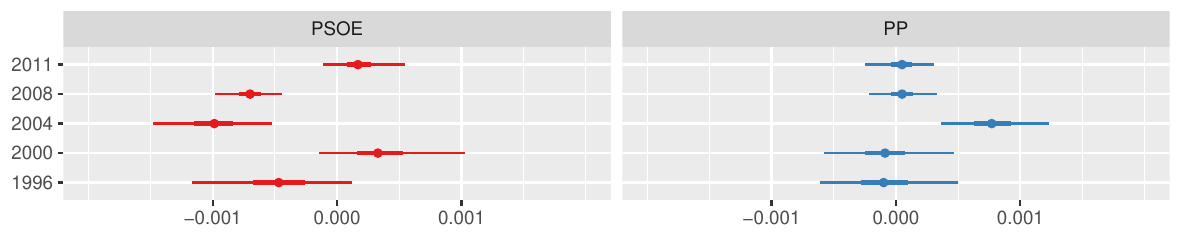}
 \caption{$\boldsymbol{\epsilon}_{t}$ marginal distributions (election trend): median (point), 50 percent credibility interval (thick line) and 95 percent credibility interval (thin line). Positive values imply that polls are trending down. \label{fig:election_trend}}
\end{figure}

Figures \ref{fig:polls_bias}, \ref{fig:election_effect} and \ref{fig:election_trend} depict the marginal distribution of pollster bias, election bias and election trend respectively\footnote{Hyperpriors are set in accordance with Stan reference priors. We sample from the models using \emph{Stan}, running 4 chains with 2000 iterations each, the first half of which we discard.}. Estimated pollster biases $\boldsymbol{\beta}$ are generally consistent with political expedience. For example, the pollster \emph{Sigma dos}, which mostly provides polls for the right-leaning newspaper \emph{El Mundo}, has a consistent bias in favour of the Popular Party. \emph{Invymark}, the pollster selected by left-leaning TV station La Sexta, shows a consistent bias in favour of the Socialist party. By contrast, the polls run by the \emph{CIS}, the public pollster, do not show any specific party bias. Election biases $\boldsymbol{\delta}_{t}$ are large, with pollsters collectively missing the PSOE-PP differential by 7 percentage points, calling into question the quality of Spanish polling and the predictability of Spanish elections in general. Estimated trend effects $\boldsymbol{\epsilon}_{t}$ are large in many elections, which confirms that some trend adjustment is necessary even within the last 30 days. Finally, election biases seem to coincide in sign and magnitude with trends, especially in the 2004 elections, but we deem our sample to be too small to draw further conclusions.

\subsection{Synthesised predictions}
\label{sec:spanish-synth}

\begin{figure}[htbp]
\centering
\includegraphics[width=\textwidth]{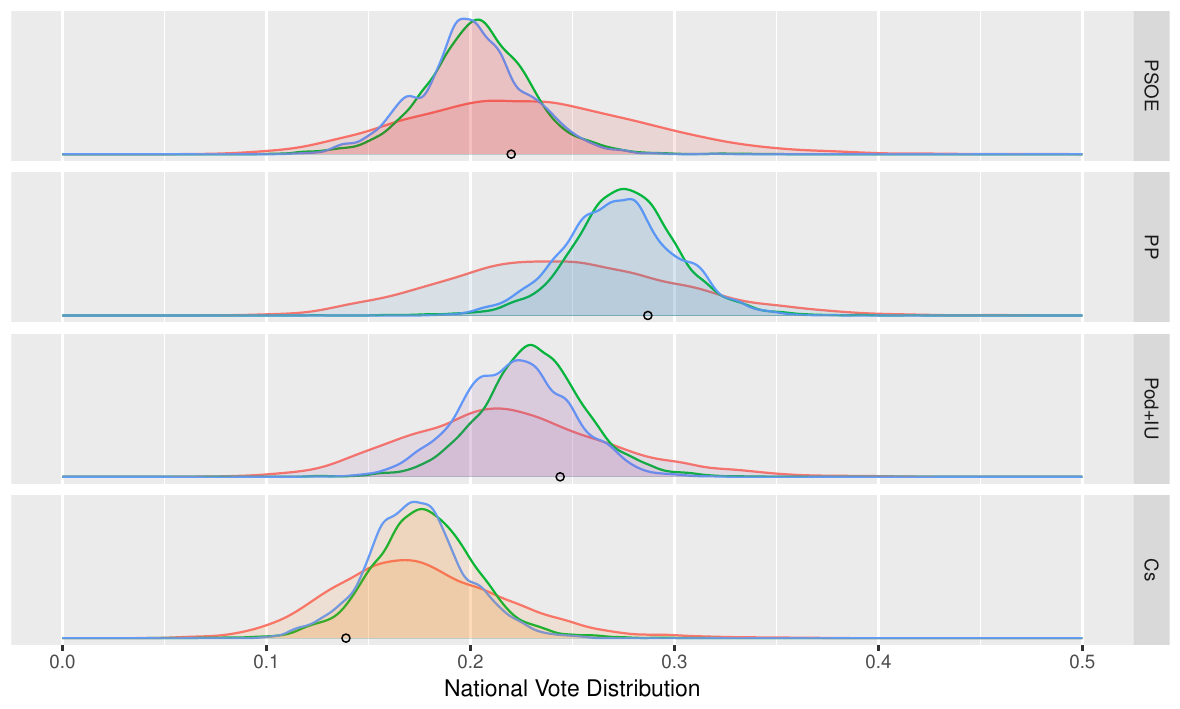}
\caption{Predictive national vote distribution: fundamental model
  (red), polls model (green), synthesis (blue). The dots represent the election result. \label{fig:synthesis}}
\end{figure}

Figure \ref{fig:synthesis} shows how the synthesis operates in the 2015 election. As in Bayesian updating for normal distributions, the posterior's location is a compromise between prior and likelihood while inverse variance is approximately additive. Aggregating polls substantially improves the PP and Podemos prediction even though the C's forecast benefits less. The benefits of aggregation are limited when pollsters show correlated errors due to herding or common methodological shortcomings. Since our framework explicitly allows for such a scenario, we manage to avoid undue confidence and preserve some probability mass at the outcome.  We observe that the location of the predictive distribution over the national vote is largely driven by the polls model. While it is difficult to obtain quantitative importance weights for prior and likelihood in general Bayesian models, such weights exist for the case where both are Gaussian. When approximating prior and posterior by Gaussian distributions, we find a weight of 35\% for the fundamental model in forming synthesised beliefs for the national average.

\begin{figure}[htbp]
\centering
\includegraphics[width=\textwidth]{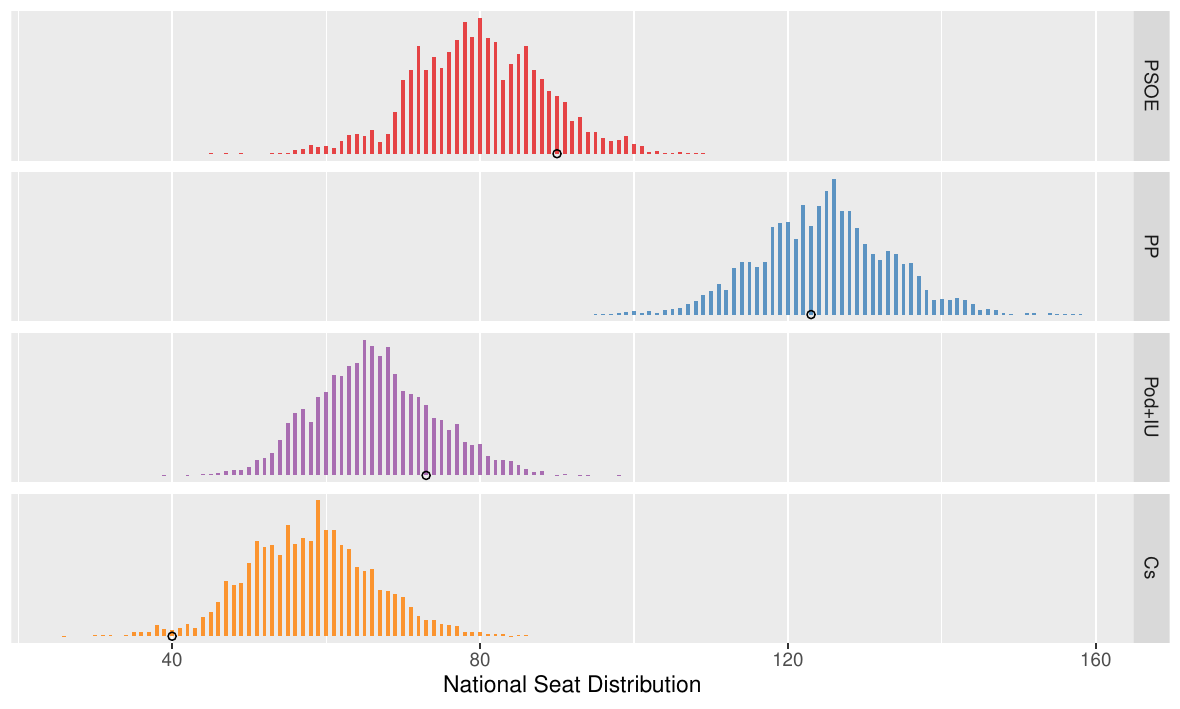}
\caption{Predictive seat distribution and election result (black dot). \label{fig:seats}}
\end{figure}

Figure \ref{fig:seats} shows the predictive seat distribution for the largest five parties. The result is close to the predictive mode for PSOE, PP and Podemos, while the result of C's is in the left tail of the predictive distribution. The predictive distribution that we generated the figure from may also be used to evaluate the probability of other events on said distribution. Examples include the probability of a party coming in first or the probability of a coalition of parties achieving a parliamentary majority. The figure exhibits some features that illustrate the benefits of our strategy to model seat allocation explicitly:

\begin{itemize}
\item PSOE and PP are granted more seats than their national vote forecast implies. This is due to the rural bias of the provincial seat allocation.
\item There are long right tails in the marginals for Podemos and C's even though national vote forecasts are symmetric. These are a consequence of the D'Hondt allocation process that delivers increasing benefits to scale.
\item The seat forecasts for PSOE and PP are wider than they are for Podemos and C's even though uncertainty when predicting the national vote is similar.
\end{itemize}

\begin{figure}
\centering
\includegraphics[width=\textwidth]{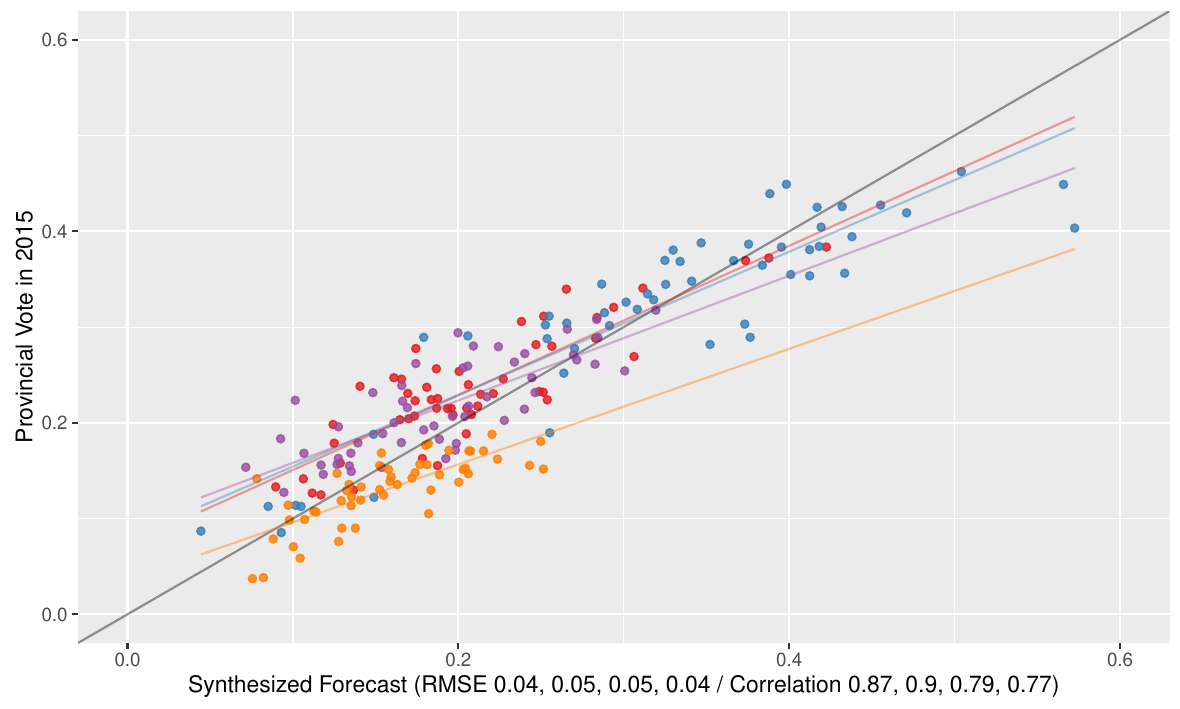}
\caption{Scatterplot of point predictions vs. outcomes and regression line. Statistics are listed in the usual party order. MSE is computed as the average squared difference between the mean prediction and the result over provinces. Legend: Legend: PSOE (red), PP (blue), Podemos+IU (purple), C's (orange). \label{fig:province}}
\end{figure}

Figure \ref{fig:province} shows point predictions versus actual results and is directly comparable to Figure \ref{fig:poststrat}. Both figures reveal that the CIS survey data is miscalibrated, i.e. it predicts a variance between provinces that is too large. The phenomenon applies to all parties and it is visible before and after post-stratification. Predictions could be recalibrated by shrinking all forecasts towards a party's national mean, but this would require that the phenomenon persists between election. Our analysis of the \emph{electoral barometers} published by CIS every trimester confirms that the miscalibration is persistent, but unfortunately the extent of the phenomenon seems to vary from one survey to the next.

\section{Predictive evaluation against alternative models}
\label{sec:eval}

While the primary appeal of our model lies in its ability to flexibly incorporate polling data into a coherent spatio-temporal probabilistic forecast, we also intended to deliver an improved point forecast relative to more straightforward approaches. In this section we elaborate several alternative models for each component as well as the hybrid model, and we show the gains in predictive accuracy achieved through our methodology. In keeping with the spirit of the paper, we also evaluate separately the two components of our hybrid model, the fundamental and the polls model, in isolation against their respective alternative models. Since many alternative models are limited to giving a point prediction of the national-level result of the parties, we only evaluate our model based on such point predictions, keeping in mind that our main object of interest is the distribution of seats. Additionally, in order to provide a fair comparison with other models, we only consider the prediction of the two traditional parties, avoiding emergent parties. However, we should notice that our modelling strategy is mostly informed by the need to accommodate the participation of new parties in political elections.

\subsection{Selecting alternative models}
\label{ssec:eval-construct}

As discussed in Section 1, the prediction of elections usually rests on the specification of a fundamental model or the aggregation of the results of electoral polls and other sources of information. Recent literature on election forecasting also resorts to some combination of prediction from fundamental variables and averages of polls. In this section we describe some popular election forecasting models and compare their predictive performance with our proposal.

We can specify an alternative fundamental model as a regression model that predicts the party's result from the growth of GDP per capita during the preceding year and its result in the previous election\footnote{This specification is similar to \cite{lewis2} and \cite{lewis3} but uses the results in the previous election instead of the government approval rate, since we want to show the predictive ability of our model not only for the incumbent but also for the main challenging party.}:
\begin{equation}
  \text{result} = \beta_{0} + \beta_{1} \times \text{lagged result} + \beta_{2} \times \text{gdp growth} + \text{residual}
\end{equation}
We consider separate versions of that model for predicting the national vote share and the log number of parliamentary seats won by the incumbent party.

We construct an alternative specification for the polls' model as a linear regression model that predicts the proportion of votes for a party as a simple average of all national level polls published up to 30 days before election day\footnote{This specification is also similar to \cite{lewis2} and \cite{lewis3} who set the predictions of the polls' model to the aggregate median vote intention. In our case the median and the mean are very similar and, since many standard models like \cite{graefe} use the mean, we decided to use the mean.}:
\begin{equation}
  \text{result} = \gamma_{0} + \gamma_{1} \times \text{polls average} + \text{residual}
\end{equation}
Finally, the alternative hybrid model combines the predictors of the fundamental model and the polls in a single regression following \cite{lewis2} and \cite{lewis3}:
\begin{multline}
  \text{result} = \delta_{0} + \delta_{1} \times \text{lagged result} + \delta_{2} \times \text{gdp growth} \\
  + \delta_{3} \times \text{polls average} + \text{residual}
\end{multline}
Analogously, as we already pointed out, we define a model for the main challenging party, which is either PP or PSOE in the set of available elections.

\subsection{Estimating the alternative models}
\label{ssec:eval-estimate}

We estimate the parameters of the alternative models through ordinary least squares (OLS).  The following numbers pertain to the incumbent model trained for predicting the 2015 election, i.e. the model that includes all elections up to 2015 in its training set. This matches the training set that we used to train our proposed model. For the fundamental model, we obtain the following parameter estimates:
\begin{gather}
  \text{votes} = 0.306 - 0.054 \times \text{lagged votes} + 0.035 \times \text{gdp growth} + \text{residual} \\
  \log \text{seats} = 5.994 - 0.255 \times \log \text{lagged seats} + 0.108 \times \text{gdp growth} + \text{residual}
\end{gather}
For the polls model, we obtain the following parameter estimates:
\begin{gather}
  \text{votes} = 0.023 + 0.943 \times \text{polls average} + \text{residual} \\
  \log \text{seats} = 3.846 + 3.007 \times \text{polls average} + \text{residual}
\end{gather}
For the synthetic model, we obtain the following parameter estimates:
\begin{multline}
  \text{votes} = 0.429 - 0.783 \times \text{lagged votes} + 0.011 \times \text{gdp growth} \\
  + 0.649 \times \text{polls average} + \text{residual}
\end{multline}
\begin{multline}
  \log \text{seats} = 8.058 - 0.743 \times \log \text{lagged seats} + 0.062 \times \text{gdp growth} \\
  + 1.468 \times \text{polls average} + \text{residual}
\end{multline}

\begin{table}
\centering
\begin{tabular}{cccrrrr}
\toprule
\multicolumn{3}{c}{}                  & \multicolumn{2}{c}{Alternative} & \multicolumn{2}{c}{Our model}  \\
\cmidrule(lr){4-5} \cmidrule(lr){6-7}
Election      & Party       & Outcome & Estimate & Residual           & Estimate & Residual           \\
\midrule
\textbf{2015} & \psoe{PSOE} & 0.220   & 0.471    & -0.251             & 0.228    & 0.008              \\
\textbf{2015} & \pp{PP}     & 0.287   & 0.331    & -0.044             & 0.244    & 0.042              \\
\textbf{2016} & \psoe{PSOE} & 0.226   & 0.143    & 0.082              & 0.223    & 0.003              \\
\textbf{2016} & \pp{PP}     & 0.330   & 0.443    & -0.112             & 0.262    & 0.067              \\
\bottomrule
\end{tabular}
\caption{Predictive accuracy of the proposed \emph{fundamental} \emph{vote} forecasting model compared to the benchmark model. Predictions are made out of sample. Both models use the same set of elections for training.\label{tb:benchmark-fundamental}}
\end{table}

Initially, we check the performance of the two components of the hybrid model. First of all, as shown in Table \ref{tb:benchmark-fundamental}, our \emph{fundamental} model outperforms the alternative fundamental model, giving errors of .042/.008 compared to .044/.251 in 2015 for the incumbent and the main challenging party (PP/PSOE). For the 2016 election our model delivers even smaller errors than the alternative model. We should keep in mind that the primary goal of the fundamental model is to provide local information on party strength, and therefore the national level point estimates are of secondary importance. See Table \ref{tb:benchmark-fundamental} for all the estimates and outcomes.

\begin{table}
\centering
\begin{tabular}{cccrrrr}
\toprule
\multicolumn{3}{c}{}                  & \multicolumn{2}{c}{Alternative} & \multicolumn{2}{c}{Our model}  \\
\cmidrule(lr){4-5} \cmidrule(lr){6-7}
Election      & Party       & Outcome & Estimate & Residual           & Estimate & Residual           \\
\midrule
\textbf{2015} & \psoe{PSOE} & 0.220   & 0.277    & -0.057             & 0.213    & 0.007              \\
\textbf{2015} & \pp{PP}     & 0.287   & 0.281    & 0.005              & 0.293    & -0.006             \\
\textbf{2016} & \psoe{PSOE} & 0.226   & 0.229    & 0.002              & 0.216    & 0.010              \\
\textbf{2016} & \pp{PP}     & 0.330   & 0.303    & 0.027              & 0.302    & 0.028              \\
\bottomrule
\end{tabular}
\caption{Predictive accuracy of the proposed \emph{polls} \emph{vote} forecasting model compared to the benchmark model. Predictions are made out of sample. Both models use the same set of elections for training.\label{tb:benchmark-polls}}
\end{table}

Secondly, the alternative \emph{polls'} model has an error of .005/.057. in 2015, whereas our pollster model has an error of .006/.007 (PP/PSOE). Thus, the quality of point estimates is slightly superior in our model. Simple polls averages predict similarly well with errors of .012/.009. See Table \ref{tb:benchmark-polls} for estimates and outcomes\footnote{Similar conclusions apply for the 2016 elections.}. In this particular case it seems that modelling the biases of the polls does not provide a large advantage over other methods but, in general, it could substantially improve the forecasting. For instance, the simple average would have been inadequate in elections where polls exhibit strong trending during the last month. Our proposed polls model accounts for such trending and should yield better point predictions in those situations.

\begin{table}
\centering
\begin{tabular}{cccrrrr}
\toprule
\multicolumn{3}{c}{}                  & \multicolumn{2}{c}{Alternative} & \multicolumn{2}{c}{Our model}  \\
\cmidrule(lr){4-5} \cmidrule(lr){6-7}
Election      & Party       & Outcome & Estimate & Residual           & Estimate & Residual           \\
\midrule
\textbf{2015} & \psoe{PSOE} & 0.220   & 0.607    & -0.387             & 0.203    & 0.017              \\
\textbf{2015} & \pp{PP}     & 0.287   & 0.273    & 0.013              & 0.275    & 0.012              \\
\textbf{2016} & \psoe{PSOE} & 0.226   & 0.261    & -0.035             & 0.212    & 0.014              \\
\textbf{2016} & \pp{PP}     & 0.330   & 0.392    & -0.062             & 0.293    & 0.037              \\
\bottomrule
\end{tabular}
\caption{Predictive accuracy of the proposed \emph{hybrid} \emph{vote} forecasting model compared to the benchmark model. Predictions are made out of sample. Both models use the same set of elections for training.\label{tb:benchmark-hybrid}}
\end{table}

The results of the comparison of the predictive performance of our model and the alternative hybrid model are presented in Table \ref{tb:benchmark-hybrid} (proportion of votes) and Table \ref{tb:benchmark-hybrid-seats} (seats). Our \emph{hybrid} model outperforms the alternative model in the 2015 and the 2016 election with respect to predicting the national vote share. The comparison of the forecast with respect to our pollster model leads to less conclusive results. In fact, in this case, our pollster model seems to work a bit better than the hybrid model in predicting the national vote. However, as we have already argued in previous sections, the most likely advantage of our methodology is related with the forecasting of parliamentary seats. Table \ref{tb:benchmark-hybrid-seats} shows that our forecast using the hybrid model reduces the error down to 2/14 seats from about 17/47 (PP/PSOE) of the alternative hybrid model in the 2015 elections\footnote{The results of our seat allocation forecasting model are also better than the ones produced by the alternative model in the 2016 election.}. The large improvement in the performance of our model in the prediction of the seats in the parliament is consistent with the objectives of the specification of our fundamental model. As we discussed previously, our fundamental model tries to get information on the geographical distribution of votes, which greatly improves the prediction of seats given the very nonlinear nature of the relationship between national proportion of votes and seat allocation.

\begin{table}
\centering
\begin{tabular}{ccrrrrr}
\toprule
\multicolumn{3}{c}{}                  & \multicolumn{2}{c}{Alternative} & \multicolumn{2}{c}{Our model}  \\
\cmidrule(lr){4-5} \cmidrule(lr){6-7}
Election      & Party       & Outcome & Estimate & Residual           & Estimate & Residual           \\
\midrule
\textbf{2015} & \psoe{PSOE} & 90.00   & 137.57   & -47.57             & 75.47    & 14.53              \\
\textbf{2015} & \pp{PP}     & 123.00  & 105.65   & 17.34              & 125.32   & -2.31              \\
\textbf{2016} & \psoe{PSOE} & 85.00   & 112.97   & -27.97             & 79.72    & 5.28               \\
\textbf{2016} & \pp{PP}     & 137.00  & 165.46   & -28.46             & 119.11   & 17.88              \\
\bottomrule
\end{tabular}
\caption{Predictive accuracy of the proposed \emph{hybrid} \emph{seats} forecasting model compared to the benchmark model. Predictions are made out of sample. Both models use the same set of elections for training.\label{tb:benchmark-hybrid-seats}}
\end{table}

\section{Conclusions}

This paper proposed a methodology to forecast electoral outcomes using the result of the combination of a fundamental model and a model-based aggregation of polls. We propose a Bayesian hierarchical structure for the fundamental model that synthesises data at the provincial, regional and national level. We use a Bayesian strategy to combine the fundamental model with the information coming for recent polls. This model can naturally be updated every time new information, for instance a new poll, becomes available. This methodology is well suited to deal with increasingly frequent situations in which new political parties enter an electoral competition, although our approach is general enough to accommodate any other electoral situation. We illustrate the advantages of our method using the 2015 Spanish Congressional Election in which two new parties ended up receiving 30\% of the votes. We compare the predictive performance of our model versus alternative models. In general the predictions of our model outperform the alternative specifications, including hybrid models that combine fundamental and polls models. Our predictions are, in relative terms, particularly accurate in predicting the seats obtained by each political party.

\bibliography{references}{}

\begin{thebibliography}{}

\bibitem[Balaguer-Coll et~al., 2015]{balaguer}
Balaguer-Coll, M.~T., Brun-Martos, M.~I., Anabel~Forte, A., and Tortosa-Ausina,
  E. (2015).
\newblock Local governments' re-election and its determinants: New evidence
  based on a bayesian approach.
\newblock {\em European Journal of Political Economy}, 39:94--108.

\bibitem[Campbell, 1992]{camp2}
Campbell, J.~E. (1992).
\newblock Forecasting the presidential vote in the states.
\newblock {\em American Journal of Political Science}, pages 386--407.

\bibitem[Campbell, 2008]{camp1}
Campbell, J.~E. (2008).
\newblock Evaluating us presidential election forecasts and forecasting
  equations.
\newblock {\em International Journal of Forecasting}, 24(2):259--271.

\bibitem[Carpenter et~al., 2017]{stan}
Carpenter, B., Gelman, A., Hoffman, M.~D., Lee, D., Goodrich, B., Betancourt,
  M., Brubaker, M., Guo, J., Li, P., and Riddell, A. (2017).
\newblock Stan: A probabilistic programming language.
\newblock {\em Journal of statistical software}, 76(1).

\bibitem[Cook and Ni, 2005]{cook}
Cook, D. and Ni, L. (2005).
\newblock Sufficient dimension reduction via inverse regression: A minimum
  discrepancy approach,.
\newblock {\em Journal of the American Statistical Association},
  100(470):410--428.

\bibitem[Dennison and Pardijs, 2016]{dennis}
Dennison, S. and Pardijs, D. (2016).
\newblock {\em The World According to Europe's Insurgent Parties}.
\newblock European Council on Foreign Relations.

\bibitem[Elinder, 2010]{elinder}
Elinder, M. (2010).
\newblock Local economies and general elections: The influence of municipal and
  regional economic conditions on voting in sweden 1985–2002.
\newblock {\em European Journal of Political Economy}, 26, Issue 2:279--292.

\bibitem[Fair, 2009]{fair}
Fair, R.~C. (2009).
\newblock Presidential and congressional vote-share equations.
\newblock {\em American Journal of Political Science}, 53(1):55--72.

\bibitem[Gayo-Avello, 2012]{gayo}
Gayo-Avello, D. (2012).
\newblock No, you cannot predict elections with twitter.
\newblock {\em IEEE Internet Computing}, 16(6):91--94.

\bibitem[Gelman et~al., 2006]{gelman2}
Gelman, A. et~al. (2006).
\newblock Prior distributions for variance parameters in hierarchical models
  (comment on article by browne and draper).
\newblock {\em Bayesian analysis}, 1(3):515--534.

\bibitem[Gelman and Hill, 2007]{gelmanbook}
Gelman, A. and Hill, J. (2007).
\newblock {\em Data analysis using regression and multilevelhierarchical
  models}, volume~1.
\newblock Cambridge University Press New York, NY, USA.

\bibitem[Gelman and King, 1993]{gelman1}
Gelman, A. and King, G. (1993).
\newblock Why are american presidential election campaign polls so variable
  when votes are so predictable?
\newblock {\em British Journal of Political Science}, 23(4):409--451.

\bibitem[Graefe, 2015]{graefe}
Graefe, A. (2015).
\newblock German election forecasting. comparing and combining methods for
  2013.
\newblock {\em German Politics}, 24(2):195--204.

\bibitem[Graefe et~al., 2014]{graefe2}
Graefe, A., Armstrong, J.~S., Jones, R.~J., and Cuzán, A.~G. (2014).
\newblock Combining forecasts: an application to elections.
\newblock {\em International Journal of Forecasting}, 30(1):43--54.

\bibitem[Hibbs, 2008]{hibbs}
Hibbs, D.~A. (2008).
\newblock Implications of the ‘bread and peace’model for the 2008 us
  presidential election.
\newblock {\em Public Choice}, 137(1):1--10.

\bibitem[Hummel and Rothschild, 2014]{hummel}
Hummel, P. and Rothschild, D. (2014).
\newblock Fundamental models for forecasting elections at the state level.
\newblock {\em Electoral Studies}, 35:123--139.

\bibitem[Klarner, 2008]{klarner}
Klarner, C. (2008).
\newblock Forecasting the 2008 us house, senate and presidential elections at
  the district and state level.
\newblock {\em PS: Political Science \& Politics}, 41(4):723--728.

\bibitem[Lewis-Beck, 2005]{lewis}
Lewis-Beck, M. (2005).
\newblock Election forecasting: Principles and practice.
\newblock {\em The British Journal of Politics \& International Relations},
  7(2):145--164.

\bibitem[Lewis-Beck and Dassonneville, 2016]{lewis2}
Lewis-Beck, M. and Dassonneville, R. (2016).
\newblock Forecasting methods in europe: synthetic models.
\newblock {\em Research and Politics}, 2(1):1--11.

\bibitem[Lewis-Beck et~al., 2016]{lewis3}
Lewis-Beck, M., Nadeau, R., and Belanger, E. (2016).
\newblock The british general election: synthetic forecast.
\newblock {\em Electoral Studies}, 41(1):264--268.

\bibitem[Linzer, 2013]{linzer}
Linzer, D.~A. (2013).
\newblock Dynamic bayesian forecasting of presidential elections in the states.
\newblock {\em Journal of the American Statistical Association},
  108(501):124--134.

\bibitem[Liu, 2001]{liu}
Liu, J. (2001).
\newblock Monte carlo strategies in statistical computing.

\bibitem[Lock and Gelman, 2010]{lock}
Lock, K. and Gelman, A. (2010).
\newblock Bayesian combination of state polls and election forecasts.
\newblock {\em Political Analysis}, 18(3):337--348.

\bibitem[Montalvo et~al., 2016]{montalvo2}
Montalvo, J., Papaspiliopoulos, O., and Stumpf-Fetizon, T. (2016).
\newblock Predicting elections with emerging political parties,.
\newblock {\em arXiv:1612.03073}.

\bibitem[Montalvo, 2011]{mont}
Montalvo, J.~G. (2011).
\newblock Voting after bombings: a natural experiment on the effect of
  terrorist attacks on democratic elections,.
\newblock {\em Review of Economics and Statistics}, 93(4):1146--1154.

\bibitem[Murthy, 2015]{murthy}
Murthy, D. (2015).
\newblock Twitter and elections: are tweets, predictive, reactive, or a form of
  buzz?
\newblock {\em Information, Communication \& Society}, 18(7):816--831.

\bibitem[Park et~al., 2004]{park}
Park, D.~K., Gelman, A., and Bafumi, J. (2004).
\newblock Bayesian multilevel estimation with poststratification: state-level
  estimates from national polls.
\newblock {\em Political Analysis}, 12(4):375--385.

\bibitem[Shirani-Merh et~al., 2018]{shirani}
Shirani-Merh, H., Rothschild, D., Goel, S., and Gelman, A. (2018).
\newblock Disentangling bias and variance in election polls.
\newblock {\em Journal of the American Statistical Association},
  113(522):607--614.

\bibitem[Silver, 2017]{silver}
Silver, N. (2017).
\newblock Fivethirtyeight’s pollster ratings,.
\newblock {\em URL https://projects. fivethirtyeight.com/pollster-ratings/}.

\bibitem[Smith and Gelfand, 1992]{smith}
Smith, A. and Gelfand, A. (1992).
\newblock Bayesian statistics without tears: A sampling-resampling
  perspective,.
\newblock {\em The American Statistician}, 46(2):84--88.

\bibitem[Stegmueller, 2013]{steg}
Stegmueller, D. (2013).
\newblock How many countries for multilevel modeling? a comparison of
  frequentist and bayesian approaches.
\newblock {\em American Journal of Political Science}, 57(3):748--761.

\bibitem[Udina and Delicado, 2005]{udina}
Udina, F. and Delicado, P. (2005).
\newblock Estimating parliamentary composition through electoral polls.
\newblock {\em Journal of the Royal Statistical Society: Series A (Statistics
  in Society)}, 168(2):387--399.

\bibitem[Voia and Ferris, 2013]{canada}
Voia, M.-C. and Ferris, J.~S. (2013).
\newblock Do business cycle peaks predict election calls in canada?
\newblock {\em European Journal of Political Economy}, 29:102--118.

\bibitem[Wang et~al., 2015]{wang}
Wang, W., Rothschild, D., Goel, S., and Gelman, A. (2015).
\newblock Forecasting elections with non-representative polls.
\newblock {\em International Journal of Forecasting}, 31(3):980--991.

\end{thebibliography}
\bibliographystyle{apalike}

\newpage

\newpage

\newpage
\appendix

\section{Appendix}
\subsection{Data}
\label{a:data}

The data on the election outcomes come from the Spanish Department of Home Affairs. For the fundamental model, we use the 2015 CIS (CIS is the Spanish National Center for Sociological Research) pre-electorals (CIS study number 3117). The study is openly available on \url{http://www.cis.es/} and includes 17452 respondents. Data was collected from October 27th to November 16th, 2015.

\begin{table}[h!]
\renewcommand\thetable{A.1}
\centering
\begin{tabular*}{\textwidth}{@{\extracolsep{\fill} } lll}
\toprule
Factor & Code & Levels \\
\midrule
Voting Intention & 1 & PSOE \\
& 2 & PP \\
& 3 & Podemos, En Com\'u Podem, En Marea, IU \\
& 4 & Ciudadanos \\
& 5 & Others \\
\midrule
Province & 1-52 & INE Province Code \\
\midrule
Municipality Population & 1 & less than 2000 inh. \\
& 2 & between 2000 and 10000 inh. \\
& 3 & more than 10000 inh. \\
\midrule
Gender & 1 & Male \\
& 2 & Female \\
\midrule
Age & 1 & 18 to 35 y.o. \\
& 2 & 36 to 55 y.o. \\
& 3 & more than 56 y.o. \\
\midrule
Education & 1 & Primary or less \\
& 2 & Secondary \\
& 3 & Tertiary \\
\midrule
Activity & 1 & Employed \\
& 2 & Unemployed \\
& 3 & Out of the labour force \\
\bottomrule
\end{tabular*}
\caption{Factors used in the fundamental model and their categories. These categorical features define 8424 distinct strata, or 162 distinct strata per province\label{tb:factors}}
\end{table}

To train the polls model, we use 157 polls published within 30 days of the 1996, 2000, 2004, 2008 and 2011 Congressional Elections. Furthermore, to generate predictions, we use 51 polls published within 30 days of the 2015 Congressional Election.

\subsection{Hierarchical modelling notation}
\label{a:notation}

Hierarchical modelling notation is a convenient way of describing models that include a lot of categorical variables as regressors. Our hierarchical modelling notation follows the standard set by \cite{gelmanbook} in \emph{Data analysis using regression and multilevel/hierarchical models}.

Consider this brief explanation of the notation. Let $\{1,\dots,I\}$ index a set of observations and $\{1,\dots,J\}$ be the indices of the levels of a categorical factor. Then, the notation $j[i]$ refers to a map $\{1,\dots,I\} \mapsto \{1,\dots,J\}$ which links each observation to its respective factor level. For instance, if the factor is gender, male has index 1, female index 2 and observation 1 is female, then $j[1] = 2$.

If $\boldsymbol{\beta}$ is the vector of coefficients pertaining to the levels of some factor, we can use hierarchical modelling notation to retrieve components of that vector. In keeping with our example, $\beta_{j[1]} = \beta_{2}$ is the coefficient of the gender of observation 1, which is equivalently the coefficient of the female level of the gender factor.

We may express this equivalently using dummy variables, but hierarchical modelling notation tends to be more concise. For example, consider a simple regression model with one categorical factor. In dummy notation, we write $y_{i} = \beta_{0} + \sum_{j} \beta_{j} x_{ij} + \epsilon_{i}$. In hierarchical modelling notation, we just write $y_{i} = \beta_{0} + \beta_{j[i]} + \epsilon_{i}$.

\subsection{The fundamental model}
\label{a:fundamental}

In the following section, let $\boldsymbol{\theta}$ refer to the set of unknowns. We model the survey response counts $\boldsymbol{s}_{n} \in \mathbf{N}^{L}$ of a stratum $n$ through a multinomial logit model:
\begin{gather}
  \boldsymbol{s}_{n} | \boldsymbol{\theta} \sim \operatorname{Multinomial}(\boldsymbol{\mu}_{n}(\boldsymbol{\theta})) \\
  \boldsymbol{\mu}_{n}(\boldsymbol{\theta}) = \operatorname{softmax}\left[\boldsymbol{\alpha} + \sum_{k} \boldsymbol{\beta}_{(k, j_{k}[i])}\right]
\end{gather}
where $k$ indexes the factors through which we define the strata (e.g. location, gender, education) and $j_{k}$ indexes the possible levels of factor $k$ (e.g. male, female, unreported for factor gender). Thus, $\boldsymbol{\beta}_{(k, j_{k})}$ is the coefficient pertaining to factor $k$ and level $j_{k}$, and $j_{k}[n]$ is the level of factor $k$ that corresponds to stratum $n$. The operator $\operatorname{softmax}$ is the multivariate generalisation of the logistic function.

We pool each factor's levels to a common prior:
\begin{equation}
  \boldsymbol{\alpha} \sim \operatorname{N}(\boldsymbol{0}, \mathbf{I}), \qquad \boldsymbol{\beta}_{(k, j_{k})} | \boldsymbol{\sigma}_{k} \sim \operatorname{N}(0, (\operatorname{diag} \boldsymbol{\sigma}_{k})^2), \qquad \boldsymbol{\sigma}_{k} \overset{\mathrm{iid}}{\sim} \operatorname{half-N}(1)
\end{equation}
While the coefficients are identified due to the prior, we stick to the standard identifiability constraint of setting all coefficients of the residual party to zero. Then, coefficients may be interpreted as changing the response probabilities relative to the residual party.

\subsection{The polls model}
\label{a:polls}

In the following section, let $\boldsymbol{\psi}$ refer to the set of unknowns except for $\boldsymbol{v}_{t}$, i.e. the result of the $t$-th election. All vectors have dimension equal to the number of parties minus 1. Dropping the last dimension is necessary to ensure that the distribution is non-degenerate. We assume that polls are generated by the following process:
\begin{equation}
  \boldsymbol{p}_{k} | (\boldsymbol{\psi}, \boldsymbol{v}_{t[k]}) \sim \operatorname{N}(\boldsymbol{v}_{t[k]} + \boldsymbol{\gamma}_{j[k]} + \boldsymbol{\delta}_{t[k]} + d_{k} \boldsymbol{\epsilon}_{t[k]}, \mathbf{\Sigma}_{j[k]})
\end{equation}
where $\boldsymbol{\gamma}_{j}$ is the time-invariant bias of pollster $j$, $\boldsymbol{\delta}_t$ is the pollster-invariant bias in election $t$, $d_{k}$ corresponds to how many days before the election poll $k$ was published and $\boldsymbol{\epsilon}_t$ is the pollster-invariant strength of the trend in a given election. $d_{k}\boldsymbol{\epsilon}_{t[k]}$ vanishes as election day approaches, but $\boldsymbol{\delta}_t$ applies to all polls until the election.

We set the following priors on the random effects:
\begin{equation}
  \boldsymbol{\gamma}_{j} | \boldsymbol{\psi} \sim \operatorname{N}(\boldsymbol{0}, \mathbf{\Sigma}_{\gamma}), \qquad \boldsymbol{\delta}_{t} | \boldsymbol{\psi} \sim \operatorname{N}(\boldsymbol{0}, \mathbf{\Sigma}_{\delta}), \qquad \boldsymbol{\epsilon}_t | \boldsymbol{\psi} \sim \operatorname{N}(\boldsymbol{0}, \mathbf{\Sigma}_{\epsilon})
\end{equation}
If we integrate out the random effects, any two polls $k \neq k'$ have joint distribution characterised by the following mean and covariance functions:
\begin{align}
  \boldsymbol{m}(k) & = \boldsymbol{v}_{t[k]} \\
  \mathbf{C}(k, k') & = \operatorname{1}_{(t[k] = t[k'])}(\mathbf{\Sigma}_{\delta} + d_{k}d_{k'}\mathbf{\Sigma}_{\epsilon}) + \operatorname{1}_{(j[k] = j[k'])}\mathbf{\Sigma}_{\gamma} + \operatorname{1}_{(k = k')}\mathbf{\Sigma}_{j[k]}
\end{align}
Accordingly, we may express the marginal polls model as a Gaussian process:
\begin{equation}
  \boldsymbol{p}_{k} | (\boldsymbol{v}_{t[k]}, \mathbf{\Sigma}_{\gamma}, \mathbf{\Sigma}_{\delta}, \mathbf{\Sigma}_{\epsilon}, \mathbf{\Sigma}_{j[k]}) \sim \operatorname{GP}(\boldsymbol{m}(k), \mathbf{C}(k, k'))
\end{equation}
The model specified up to now defines a likelihood of polls given an upcoming election result $\boldsymbol{v}_{t^{*}}$. We may complete the specification by adding the flat prior
\begin{equation}
  p(\boldsymbol{v}_{t^{*}}) \propto 1
\end{equation}
thus allowing us to sample from the joint posterior of parameters and upcoming election result. Alternatively, we may use said likelihood to weight samples from some other prior over the upcoming election, e.g. our fundamental model.

\end{document}